\documentclass[12pt]{article} 
\usepackage{amssymb}
\begin{document} 
\begin{center}
          {\large \bf Polarization properties of the quark-gluon medium } 

\vspace{0.5cm}                   
{\bf I.M. Dremin}

\vspace{0.5cm}              
          Lebedev Physical Institute, Moscow 119991, Russia

\end{center}

\begin{abstract}
Collective properties of the quark-gluon medium induced by its polarization are
described by macroscopic QCD equations. The parton currents traversing it lead 
to emission of Cherenkov gluons, the wake effect and the transition radiation.
Comparison with experimental data of RHIC reveals large value of the 
chromopermittivity of the medium while cosmic ray data at higher energies
(close to LHC) favor much smaller values. The dispersion equations show that 
the proper modes of the medium are unstable. 
\end{abstract}

The electrodynamics of the ordinary matter is treated by introducing the
dielectric permittivity  $\epsilon $ which describes its collective response 
to external currents. An external field $\bf E$ induces the polarization 
$\bf P$ of the medium so that
\begin{equation}
{\bf P}= \frac {\epsilon - 1}{4\pi }{\bf E}.     \label{pe}
\end{equation}
This polarization is in charge of many collective effects. We'll use the 
analogy of QED and QCD to study the similar effects in heavy-ion collisions.   
 
Experiment clearly shows that the products of heavy-ion collisions can
not be described as results of independent pp-collisions. The quark-gluon 
medium formed in the collisions of high energy nuclei, surely, possesses some 
collective properties (see the review paper \cite{dl10}). The widely used 
notion of the quark-gluon plasma asks for macroscopic description of this
state of matter. Yet it is unclear how this state
is formed. Initial nuclei described as Color Glass Condensate (CGC) collide
and at the very first moment the longitudinal fields forming the strings or
tubes (Glasma) appear but soon they are destroyed by instabilities and
the folklore ascribes this stage to the formation of the quark-gluon plasma
with subsequent fast thermalization leading to the liquid described by
hydrodynamics. The chromodynamic characteristics of this state are described
by macroscopic QCD (with the chromopermittivity taking place of the dielectric
permittivity) and are measured experimentally by its response to high
energy (high $p_T$) quarks and gluons traversing it. 

Before delving in details, let us describe main effects we have in mind.
First, the in-medium QCD equations predict the emission of Cherenkov gluons 
which leads to the so-called ringlike events. The fit to experimental data of 
RHIC allows to find out both real (quite large!) and imaginary (relatively 
small!) parts of the chromopermittivity. The large real part shows the high 
density of the quark-gluon plasma while small imaginary part favors 
penetration of partons and, consequently, observation of the effect.
The transverse part of the chromopermittivity tensor is in charge of these
processes. In principle, this effect would survive even for the real 
chromopermittivity.

Another observed experimental effect due to Cherenkov 
gluons is the universal asymmetry of shapes of resonances traversing 
the quark-gluon medium as clearly seen from dilepton modes of their decays
(studied at SPS and some other accelerators).

Beside Cherenkov gluons there exists the collective classical wake effect due 
to the trail behind the penetrating parton which has been observed in experiment 
as the shift of the positions of two-hump maxima in semi-central nuclei 
collisions at RHIC. In distinction to Cherenkov gluons this effect is induced 
by the longitudinal part of the chromopermittivity tensor and, moreover, is 
proportional to its imaginary part. 

The transition radiation of gluons may also appear but we'll not describe it 
here (see \cite{dl10}).

Concerning possible new effects at LHC, one can await for Cherenkov gluons 
emitted by extremely high energy forward moving partons in non-trigger 
experiments where the values of the chromopermittivity could be very close to 1
(favored by the cosmic ray data). The dispersion law of the quark-gluon medium 
at high energies drastically differs from that in the electromagnetism. Its 
main feature is the excess of the chromopermittivity over 1. The definite model 
inspired by some experimental data about the properties of hadronic forward 
scattering amplitudes is considered. The dispersion equation shows 
that the quark-gluon medium is unstable and responses to external high energy 
partons by creation of Cherenkov gluons with specific properties. 
  
At the classical 
level QCD equations are similar to those of QED. Therefore it is quite natural 
to use the analogy with electrodynamical processes in matter. The in-medium QCD
equations differ from the in-vacuum equations by introducing a 
chromopermittivity (called here $\epsilon $ as well) of the quark-gluon medium 
(its collective response!). 

Analogously to electrodynamics, the medium is accounted for if $\bf E$ is
replaced by ${\bf D} =\epsilon {\bf E}$ in $F^{\mu \nu} $.
In terms of potentials the equations of {\it in-medium} gluodynamics are cast 
in the form \cite{inmed}
\begin{eqnarray}
\bigtriangleup {\bf A}_a-\epsilon \frac{\partial ^2{\bf A}_a}{\partial t^2}=
-{\bf j}_a -
gf_{abc}(\frac {1}{2} {\rm curl } [{\bf A}_b, {\bf A}_c]+
\epsilon \frac {\partial }
{\partial t}({\bf A}_b\Phi _c)+\frac {1}{2}[{\bf A}_b {\rm curl } {\bf A}_c]-  \nonumber \\
\epsilon \Phi _b\frac 
{\partial {\bf A}_c}{\partial t}- 
\epsilon \Phi _b {\rm grad } \Phi _c-\frac {1}{2} gf_{cmn}
[{\bf A}_b[{\bf A}_m{\bf A}_n]]+g\epsilon f_{cmn}\Phi _b{\bf A}_m\Phi _n), 
\hfill \label{f.6}
\end{eqnarray}

\begin{eqnarray}
\bigtriangleup \Phi _a-\epsilon \frac {\partial ^2 \Phi _a}
{\partial t^2}=-\frac {\rho _a}{\epsilon }+ 
gf_{abc}(-2{\bf A}_c {\rm grad }\Phi _b+{\bf A}_b
\frac {\partial {\bf A}_c}{\partial t}-\epsilon 
\frac {\partial \Phi _b}{\partial t}
\Phi _c)+  \nonumber  \\
g^2 f_{amn} f_{nlb} {\bf A}_m {\bf A}_l \Phi _b. \hfill  \label{f.7}
\end{eqnarray}

The classical equations are obtained if all terms with explicitly shown
coupling constant $g$ are omitted, and then they remind those of QED.
For the current with velocity ${\bf v}$ along the $z$-axis:
\begin{equation}
\label{f.11}
{\bf j}({\bf r},t)={\bf v}\rho ({\bf r},t)=4\pi g{\bf v}\delta({\bf r}-{\bf v}t)
\end{equation}
the classical lowest order solution of in-medium gluodynamics is \cite{inmed}
\begin{equation}
\label{f.12}
\Phi ^{(1)}({\bf r},t)=\frac {2g}{\epsilon }\frac {\theta
(vt-z-r_{\perp }\sqrt {\epsilon v^2-1})}{\sqrt {(vt-z)^2-r_{\perp} ^2
(\epsilon v^2-1)}},
\end{equation}
and
\begin{equation}
{\bf A}^{(1)}({\bf r},t)=\epsilon {\bf v} \Phi ^{(1)}({\bf r},t),   \hfill  \label{f.13}
\end{equation}
where the superscript (1) indicates the solutions of order $g$, 
$r_{\perp }=\sqrt {x^2+y^2}$ is the cylindrical coordinate; $z$ is the
symmetry axis. 

This solution describes the emission of Cherenkov gluons at the typical angle
\begin{equation}
\label{f.10}
\cos \theta = \frac {1}{v\sqrt {\epsilon }}.
\end{equation}
It is constant for constant $\epsilon >1$. Such effect was first observed
in the cosmic ray data \cite{apan, d1, d0} and called as ringlike events
(by analogy with famous Cherenkov rings in ordinary matter).
For absorbing media $\epsilon $ acquires the imaginary part. The sharp front 
edge of the shock wave (\ref{f.12}) is smoothed.  The angular distribution 
of Cherenkov radiation widens. The $\delta $-function at the angle 
(\ref{f.10}) is replaced by the a'la Breit-Wigner shape \cite {gr, dklv} 
with maximum at the same angle (but $\vert \epsilon \vert $ in place of 
$\epsilon $) and the width proportional to the imaginary part. The ringlike 
distribution of particles around the (away-side) jet traversing the 
quark-gluon medium was observed in the form of two humps when projected
on the diameter of the ring. This is completely analogous to what was done
by Cherenkov in his original publications. It has been used for fits of RHIC 
data \cite{dklv}. Both real and imaginary parts of $\epsilon = \epsilon _1 + 
i\epsilon _2$ were taken into account. For two-particle correlations measured
by STAR and PHENIX it was found that they are, correspondingly, about 6 and 0.8.

The real part of the chromopermittivity can be expressed through the real 
part of the forward scattering amplitude ${\rm Re}F_0(\omega)$ of the refracted 
quanta
\begin{equation}
\label{f.19}
{\rm Re} \Delta \epsilon = {\rm Re} \epsilon (\omega ) -1= \frac {4\pi N_s 
{\rm Re} F_0(\omega)}{\omega ^2}=\frac {N_s\sigma (\omega )\rho (\omega )}{\omega }   
\end{equation}
with
\begin{equation}
{\rm Im} F_0(\omega )=\frac {\omega }{4\pi }\sigma (\omega ).
\end{equation}
Here $\omega$ denotes the energy, $N_s $ is the density of scattering centers,
$\sigma (\omega)$ the cross section and $\rho (\omega)$ the ratio of real to 
imaginary parts of the forward scattering amplitude $F_0(\omega)$. The large 
value of ${\rm Re} \epsilon $ presented above shows that the density of the
scattering centers in the quark-gluon plasma is very high. It can be estimated
as exceeding 20 within the single proton volume.

The emission of Cherenkov gluons is possible only for processes with positive 
${\rm Re} F_0(\omega)$ or $\rho (\omega)$. The available data of
hadronic experiments show that the necessary condition for Cherenkov effects 
may be satisfied at least within two energy intervals - those of resonance
production (${\rm Re} F_0(\omega)>0 $ in left wings of all Breit-Wigner 
resonances!) and at extremely high energies (where 
${\rm Re} F_0(\omega)$ becomes positive for processes with all measured
initial particles, i.e. it has the universal character which can be related
to universality of increase of total cross sections with the collision energy).

The first region is typical for the comparatively low energies of secondary
particles registered in SPS and RHIC experiments. ${\rm Re} F_0(\omega)$ is 
always positive (i.e., $\epsilon >1$) within the low-mass wings of the 
Breit-Wigner resonances. Therefore, Cherenkov gluons can be emitted in
these energy intervals. 

The asymmetry of the $\rho $-meson mass shape observed in leptonic decays 
of $\rho $-mesons created in nuclei collisions at SPS \cite{da} was explained by 
appearence of the additional collective effect, namely that of 
emission of low-energy Cherenkov gluons \cite{dnec, drem1}
inside the left (low mass) wing of the Breit-Wigner resonance. It is predicted
that this feature should be common for all resonances traversing the nuclear
medium according to the above noted universality. Some preliminary experimental 
indications which favor this conclusion have appeared for other resonances as 
well \cite{dnec}.

The experimental data of STAR and PHENIX collaborations at RHIC on two- and 
three-particle azimuthal correlations in central collisions discussed above 
also deal with rather low energies of secondary particles. Moreover, the new 
effect of humps shift due to wake radiation was observed in 
mid-central nuclear collisions \cite{holz} and explained with the same values of
$\epsilon $ \cite{mpla}. 

At extremely high energies the properties of the quark-gluon medium may differ
strongly. Inspired by (\ref{f.19}) one can use the model with chromopermittivity
behaving above some threshold $(\omega _{thr})$ as 
\begin{equation}
{\rm Re}\epsilon =1+\frac {\omega _0^2}{\omega^2},         \label{qcd}
\end{equation}
where $\omega_0$ is some real free parameter. 
The classical equations derived from (\ref{f.6}), (\ref{f.7}) and written
in the momentum space have solution if the following dispersion equation
is valid
\begin{equation}
{\rm det}(\omega, {\bf k})=\vert k^2\delta _{ij}-k_ik_j-\omega ^2\epsilon _{ij}
\vert =0.  \label{disp}
\end{equation}
It is of the sixth order in momenta dimension. However, the sixth order terms 
cancel and (\ref{disp}) leads to two equations (of the second order):
\begin{equation}
k^2-\omega ^2-\omega _0^2=0,     \label{plas}
\end{equation}
\begin{equation}
(k^2-\omega ^2-\omega _0^2)(1+\frac {\omega_0^2}{\omega^2})-
\frac {\omega _0^4k_t^2}{\omega ^2(\omega-k_z)^2\gamma }=0.  \label{bunch}
\end{equation}
They determine the internal modes of the medium and the bunch propagation
through the medium, correspondingly.
The equation (\ref{plas}) shows that the quark-gluon medium is unstable because
there exists the branch with ${\rm Im}\omega >0$ for modes $k^2<\omega _0^2$.
Thus the universal energy increase of the hadronic total cross sections is 
directly related to the instability of the quark-gluon medium by the 
positiveness of ${\rm Re}F_0(\omega )$ at high energies.

The equation (\ref{bunch}) has solutions corresponding to Cherenkov gluons 
emitted by the impinging bunch and determined by the last term in (\ref{bunch}).
The solutions of the disperion 
equation (\ref{disp}) determine the Green function of the system 
\begin{equation}
G(t,z)=\frac {1}{2\pi^2}\int _{-\infty}^{\infty}dk\int _{C(\omega )}\frac {1}
{{\rm det}(\omega, {\bf k})}\exp (-i\omega t+ikz)d\omega,      \label{green}
\end{equation}
where the contour $C(\omega )$ passes above all singularities in the integral.
Therefore, the positive ${\rm Im}\omega $ found in solutions of (\ref{bunch})
corresponds to the absolute instability of the system. 
The instability exponent decreases asymptotically as $\gamma ^{-1/3}$ and is 
about 16 times smaller at LHC compared to RHIC.

It happens that Cherenkov gluons are emitted with constant transverse momentum 
$k_t=\omega _0$ and their number is proportional to 
$[d\omega /\omega ^2]\Theta (\omega -\omega _{thr})$ 
for $\epsilon (\omega )$ given by Eq. (\ref{qcd}). It differs from the 
traditional folklore of constant emission angle of Cherenkov radiation and the 
number of gluons $\propto d\omega $ (or the total energy loss proportional
to $\omega d\omega $) which is correct only for constant chromopermittivity. 

Thus I briefly demonstrated the statements claimed above. For more
complete version I again refer to the review paper \cite{dl10}. In particular,
quantum effects of Eqs. (\ref{f.6}), (\ref{f.7}) (e.g., the color rainbow) are 
not discussed here.

This work was supported by RFBR grants 09-02-00741; 08-02-91000-CERN and
by the RAN-CERN program.

\end{document}